**High transport critical current density above 30 K in pure Fe-clad $MgB_2$ tape**

S. Soltanian,[1] X.L. Wang,[1] I. Kusevic[2], E. Babic[2], A.H. Li,[1] H. K. Liu[1], E.W. Collings[3], and S.X. Dou[1]

[1]Institute for Superconducting and Electronic Materials, University of Wollongong, NSW 2522, Australia

[2]Physics Department, University of Zagreb, Zagreb, Croatia,

[3]Department of Materials Science and Engineering, Ohio State University, OH 43210-1179

## Abstract

Fe-clad $MgB_2$ long tapes have been fabricated using a powder-in-tube technique. An Mg + 2B mixture was used as the central conductor core and reacted in–situ to form $MgB_2$. The tapes were sintered in pure Ar at 800 $^o$ C for 1 h at ambient pressure. SEM shows a highly dense core with a large grain size of 100 μm. The Fe clad tape shows a sharp transition with transition width of $\Delta T_c$ of 0.2 K and $T_{c0}$ at 37.5 K. We have achieved the highest transport critical current reported so far at $1.6 \times 10^4$ A/cm$^2$ for both 29.5 K in 1 Tesla and 33 K in null field. R-T and critical current were also measured for fields perpendicular and parallel to the tape plane. The iron cladding shielded on the core from the applied external field, with the shielding being less effective for the field in the tape plane. Fe cladding may be advantageous for some applications as it could reduce the effects of both the self-field and external fields.

## 1. Introduction

The discovery of superconductivity at 39 K in $MgB_2$ [1] has generated great interest worldwide in both fundamental studies and practical applications. High critical current densities have been observed in bulk samples, regardless of the degree of grain alignment [2]. This would be an advantage for making wires or tapes with no degradation of $J_c$, in contrast to the degradation due to grain boundary induced weak-links which is a common and serious problem in cuprate high temperature superconductors. Critical current densities on the order of $10^4$ to $10^5$ A/cm$^2$ at 4.2 K have been reported by several groups for polycrystalline $MgB_2$ bulk samples [3-8].

A significantly large $J_c$ of $10^6$ A/cm$^2$ at 4.2 K and 1 T and substantial enhancement of the irreversibility line have been reported in high quality epitaxial $MgB_2$ thin films grown on $Al_2O_3$ and $SrTiO_3$ single crystal substrates [9, 10]. These results give further encouragement to the development of $MgB_2$ for high current applications. However, the fabrication of metal clad $MgB_2$ tapes or wires will be essential to meet the requirements of most such high current applications.

The first short wire of $MgB_2$ reported was made by exposing boron filaments to magnesium vapour [11], with a magnetic $J_c$ of $10^5$ A/cm$^2$ at 4.2K. For tape fabrication, it is very important to find a suitable sheath material for $MgB_2$ which does not degrade the superconductivity. So far, several metal clad $MgB_2$ tapes or wires have been fabricated with sheath materials such as Nb [12], Cu, Ag, and Ni [13, 14], as well as Fe. A transport $J_c$ of $10^4$ A/cm$^2$ at 4.2 K has been obtained for Ag/$MgB_2$ using reacted $MgB_2$ without any heat treatment [14]. It was found in fact, that any heat treatment gave rise to the degradation of $J_c$. A very high transport $J_c$ of $10^5$ A/cm$^2$ at 4.2K has been achieved for an unsintered Ni/$MgB_2$ tape [13]. In the case where a mixture of Mg + B powder is used, a higher $J_c$ was obtained for Cu clad $MgB_2$ tapes, but such tapes require a very long 48 h sintering at 620$^o$C [14]. The use of proper metal cladding is critical for making $MgB_2$ wires or tapes because the magnesium in $MgB_2$ tends to react with many metals such as Cu or Ag during the sintering at around 900-1000$^o$C. Fe is an inert metal which has little or no mutual solubility with Mg and thus does not form intermetallic compounds with it. Based on this fact, $MgB_2$ thick films with a $J_c$ of 8 x $10^4$ A/cm$^2$at 20 K were successfully grown on stainless steel which mainly consists of iron [15]. A high transport $J_c$ on the order of $10^4$-$10^5$ A/cm$^2$ at 20 K and 4.2 K has been reported for Cu/Fe/$MgB_2$ tapes where reacted $MgB_2$ powders were used as the core conductor and sintered at 900 $^o$C for 1 h [16].

It should be pointed that most of the transport $J_c$ results reported on $MgB_2$ wires and tapes are limited to 4.2 K so far. No high transport $J_c$ values have been reported at temperatures above 20K. To take advantages of the relatively high $T_c$ of 39 K for $MgB_2$ superconductor it is essential to have high $J_c$ values at temperatures above 20K. For example, the boiling point of hydrogen at atmospheric pressure is 20.13 K, so that it is possible to use liquid hydrogen or cold hydrogen gas as a cryogen for cooling $MgB_2$ wires. This requires that the $MgB_2$ wires have sufficiently high $J_c$ values at around 25 K. In this rapid communication, we present the fabrication of pure Fe clad $MgB_2$ tapes with a very high transport $J_c$ above $10^4$ A/cm$^2$ at 30 K and 1 Tesla and $I_c$ greater than 150 A.

## 2. Experimetal

Standard powder-in-tube methods were used for the Fe clad $MgB_2$ tape. The pure Fe tube had an outside diameter (OD) of 10 mm, a wall thickness of 1 mm, and was 10 cm long. One end of the tube was sealed, and the tube was filled in with magnesium (99% purity) and amorphous boron (99%) with the stoichiometry of $MgB_2$. The remaining end was crimped by hand. The composite was drawn to a 2-3 mm diameter rod 2 meters long, with the drawing followed by subsequent cold rolling to ribbon over many steps. Several short samples 2 cm in length were cut from the ribbon. These pieces were then sintered in a tube furnace over a temperature range from 600-1000$^o$C for 1-48h. A high purity argon gas flow was maintained throughout the sintering process. The mass loss after sintering is very small, less than 1%, the same as has been reported for Cu/Fe/$MgB_2$.

## 3. Results and Discussion

Scanning electron microscopy (SEM) photomicrographs for the Fe clad tape after sintering are shown in Fig.1. Fig. 1 (a) is a typical transverse cross section of an Fe clad tape. It clearly shows that the $MgB_2$ core presents a homogeneous cross section. Fig. 1 (b) is the longitudinal cross-sectional micrograph showing good core homogeneity. Fig. 2 represents the high magnification microstructure of the core surface after mechanically removal of the Fe sheath material. This micrograph shows a dense microstructure with a grain size of about 100 μm. Our results showing large grain sizes are very different from those seen in the reported Cu/Fe/$MgB_2$ tape which was made using reacted $MgB_2$ powders with a starting grain sizes of 3 μm. The final grain size was significantly reduced to 120 nm due to the occurrence of substantial grain refinement during the wire fabrication process [16]. In our present work, $MgB_2$ grains can easily be grown to large sizes during the high temperature sintering above 700°C for more than 1 h resulting liquid phase because the magnesium can easily react with boron to form large grains.

The critical current of the Fe clad $MgB_2$ tape was measured by the standard four-probe method. The sample used for the measurement has a length of 20.5 mm and a width of 3 mm. The $MgB_2$ core cross-section in this sample is very irregular, similar to that is shown in Fig. 1. Its average dimensions are $2.1 \times 0.45$ mm$^2$. Therefore the core cross-section is about $9.45 \times 10^{-3}$ cm$^2$. The current and voltage contacts were soldered with Wood's alloy (giving a current contact resistance lower than 10 mΩ), and the distance between the voltage contacts was 7.5 mm.

The temperature dependence of resistance (R-T) was measured using an ac current (frequency 18.4 Hz, I =1 mA). Fig. 3 shows R-T measured at zero field over a wide temperature range from 300 to 10 K. It shows a sharp transition with a transition width $\Delta T_c$ of 0.2 K and $T_{c0}$(midpoint) at 37.5 K. We also measured R-T around the transition temperature using two different geometries for different fields orientations: the field perpendicular to the tape plane (denoted with "o") and field parallel to the tape plane (denoted with "p"). In both cases, the field direction was perpendicular to the current flow direction. With these field orientations, superconducting transitions and $T_{c0}$ for the parallel field are shifted to lower temperatures compared to those with the field perpendicular to the tape plane, as shown in Fig. 3. This clearly indicates that the iron cladding effectively shields the core from the applied external field, the shielding being less effective for fields parallel to the tape plane.

Critical current measurements were made with a pulse method, with the current pulse linearly rising from zero to maximum current. The pulse duration was 20 ms for T>33 K and 10 ms for T<30 K. The voltage was amplified and recorded on a digital storage oscilloscope together with the voltage across a standard resistor, giving the current flowing through the sample. The temperature of the sample holder was monitored during the measurement with a gold-chromel thermocouple, showing a temperature rise after a pulse of approximately 0.2 K at currents higher than 150 A (The next pulse was applied after the temperature had fallen to denoted values). We estimate that the temperature rise of the sample itself was somewhat higher, but that just means that the critical current obtained is underestimated.

Although the above mentioned procedure for critical current measurements enables the determination of $I_c$ with the 1μV/cm criterion, there were two problems: 1) the current pulse causes magnetization of the iron cladding, which gives a spurious voltage signal $V_m$ superimposed on the voltage of the superconductor $V_s$. Since $V_m$ was appreciable in fields B<0.4 T, $I_c$ in these fields was determined as the current at which the overall voltage starts to increase above the decaying $V_m$. However, the error in $I_c$ determination is 10% at most, and because of heating, the real $I_c$ is probably higher than the results obtained (especially at lower fields); 2) magnetization depends on the field direction. For that reason, we measured $I_c$ for both field directions (denoted on the figures in the same way as for resistance measurements) at 35 K.

The critical currents measured at temperatures above 29 K and in fields up to 1 T are shown in Fig. 4. Since the shielding from the applied field was less for fields parallel to the tape plane than for fields perpendicular to it, the critical current at a nominal external applied field is less for fields parallel to the tape plane. $I_c$ increased from 10 to 164 A as the temperature decreased from 36.4 to 30 K, and changes of $I_c$ with field were smooth for 32 and 30 K. The critical current density $J_c$ is calculated using the calculated core cross section of $9.45 \times 10^{-3}\ \mathrm{cm}^2$, and its value is shown on the right axis of Fig. 4. We can see that the Fe clad $MgB_2$ has a very high transport $J_c$ of above $10^4$ A/cm$^2$ for fields <0.5 T at 33.2K and for fields < 0.8 T at 32 K. The highest $J_c$ is about $1.7 \times 10^4$ A/cm$^2$ at 29.5 K and 1 T. To the best of our knowledge, these values of $J_c$ are the highest reported so far. The real $I_c$ values (meaning those in the $MgB_2$ core without Fe-cladding) at these fields are probably slightly lower than those in Fe-clad $MgB_2$ due to iron shielding, meaning that fields >2.5 T are needed in order to avoid shielding effects.

Regarding the effect of Fe shielding, when there is no external field applied the transport current will generate a self field surrounding the tape. Because the Fe sheath is ferromagnetic the flux lines of the self-induced field will suck into the Fe sheath particularly at the edges of the tape. Thus, the Fe sheath will reduce the effect of the self field on the critical current. Where external fields are applied, the Fe sheath acts as shielding to reduce the effect of the external fields. This may be beneficial for some applications such as power transmission as the critical current density will be higher than for tapes without an Fe sheath.

In conclusion, Fe-clad $MgB_2$ tapes fabricated by PIT techniques and sintered in pure Ar at 800 $^{o}$C for 1 h at ambient pressure show a highly dense core with a large grain size of 100 μm. They have a sharp transition with the transition width $\Delta T_c$ of 0.2 K and $T_{c0}$(midpoint) at 37.5 K. A transport critical current of $1.6 \times 10^4$ A/cm$^2$ for both 29.5 K in 1 Tesla and for 33 K in null field has been obtained, the highest reported so far. R-T and critical current as a function of magnetic field measured for fields perpendicular and parallel to the tape plane indicate that the iron cladding shields the core from the applied external field, the shielding being less effective for fields parallel to the tape plane.

Acknowledgements

The authors thank Dr T. Silver for her helpful comments and discussion. This work is supported in part by funding from the Australian Research Council and the University of Wollongong. S. Soltanian would like to thank the Department of Physics, University of Kurdistan, Iran for providing financial support for his PhD study at the University of Wollongong.

References:

1. J. Nagamatsu, N. Nakagawa, T. Muranaka, Y. Zenitani, and J. Akimitsu. Nature, 410 (2001) 63
2. D.C. Larbalestier, M.O. Rikel, L.D. Cooley, A.A. Polyanskii, J.Y. Jiang, S. Patnaik, X.Y. Cai, D.M. Feldmann, A. Gurevich, A.A. Squitieri, M.T. Naus, C.B. Eom, E.E. Hellstrom, Nature, 410 (2001) 186
3. Y. Takano, H. Takeya, H. Fujii, H. Kumakura, T. Hatano, K. Togano, Cond-mat/01020167
4. Y. Bugoslavsky, G.K. Perkins, X. Qi, L.F. Cohen, and A.D. Caplin. , Cond-mat/0102353
5. Kim Mun-Seog, Jung C. U, Park Min-Seok, Lee S. Y, Kim Kijoon H. P, Kang W. N, and Lee Sung-Ik, Cond-mat/0102338
6. Wen H. H, Li S. L, Zhao Z. W, Ni Y. M, Ren Z. A, Che G. C, Yang H. P, Liu Z. Y, and Zhao Z. X., Cond-mat/0102436
7. S.X. Dou, X.L. Wang, J. Horvat, D. Milliken, E.W. Collings, and M.D. Sumption, cond-mat/0102320
8. Kim Kijoon H. P, Kang W. N, Kim Mun-Seog, Jung C. U, Kim Hyeong-Jin, Choi Eun-Mi, Park Min-Seok, and Lee Sung-Ik, Cond-mat/0103176
9. Kang W. N, Kim Hyeong-Jin, Choi Eun-Mi, Hung C. U, Lee Sung-Ik, cond-mat/0103179
10. C.B. Eom, et al., cond-mat/0103425, submitted to Nature
11. Canfield P. C, Finnemore D. K, Bud'ko S. L, Ostenson J. E, Lapertot G, Cunningham C. E, and Petrovic C, Phys. Rev. Lett., 86 (2001) 2423

12. Sumption M. D, Peng X, Lee E, Tomsic M, and Collings E. W., Cond-mat/0102441
13. G. Grasso, A. Malagoli, C. Ferdeghini, S. Roncallo, V. Braccini, M. R. Cimberle and A. S. Siri, Cond-mat/0103563
14. B.A. Glowacki, M. Majoros, M. Vickers, J. E. Evetts, Y. Shi and I. McDougall, Supercond. Sci. Technol., 14 (2001) 193
15. A. H. Li, X .L. Wang, M. Ionescu, S. Sotonian, J. Horvat, T. Silver, H. K. Liu, and S.X. Dou, Physica C, submitted, cond-mat/0104501
16. S. Jin, H. Mavoori and R. B. van Dover, submitted to Nature. Cond-mat/0104236

Figure Captions

Fig. 1. SEM image for a typical transverse(a) and a longitudinal (b) cross-section.

Fig. 2.High magnification microstructure of the core surface after the top Fe sheath material has been removed mechanically.

Fig.3. R-T curves for Fe/$MgB_2$ tape measured at different fields with the field perpendicular (denoted as "p") and parallel (denoted as "o") to the tape plane.

Fig. 4. Field dependence of $I_c$ (left axis) and $J_c$ (right axis) at different temperatures with fields perpendicular and parallel to the tape plane.

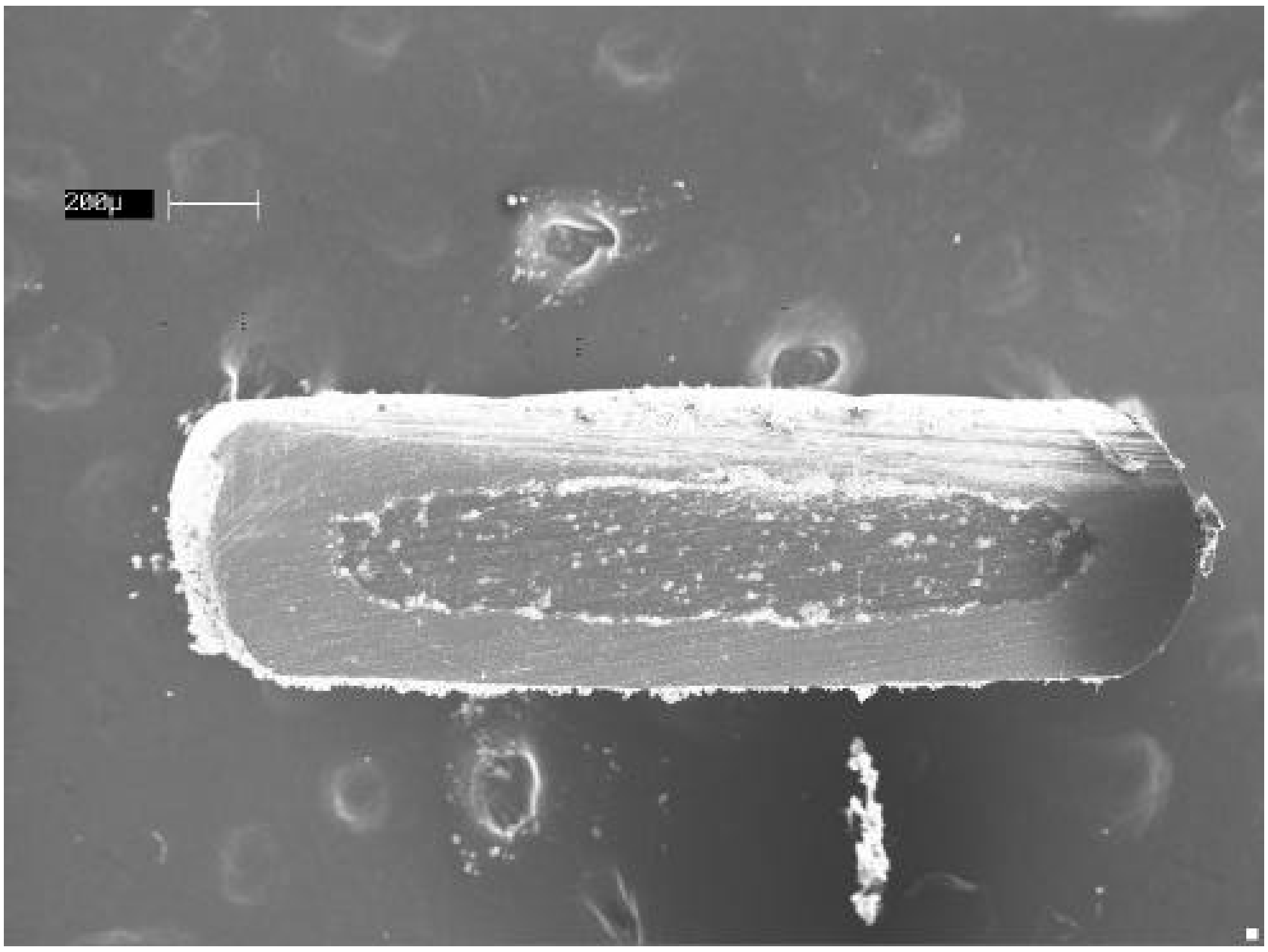

Fig. 1 (a).

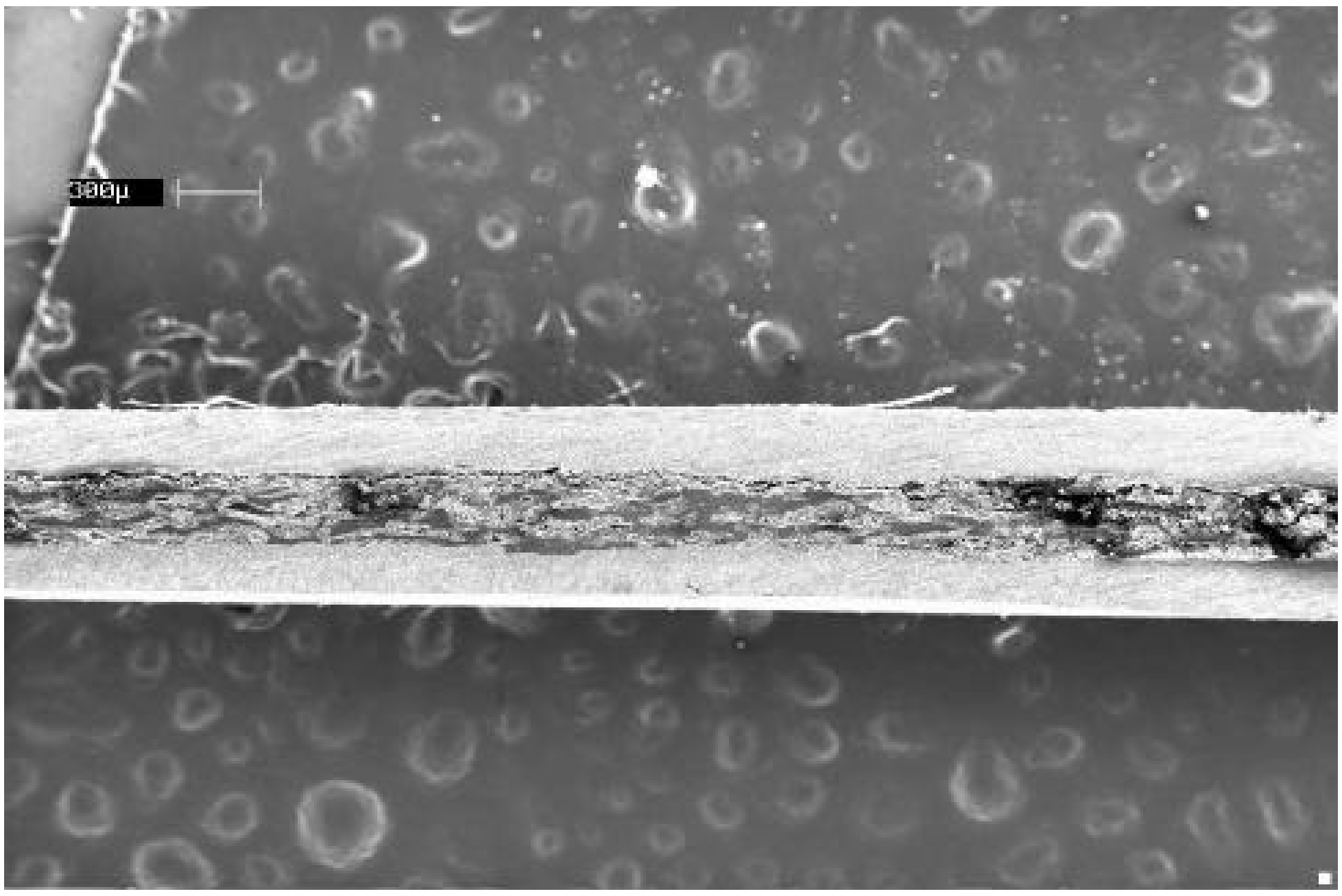

Fig.1(b).

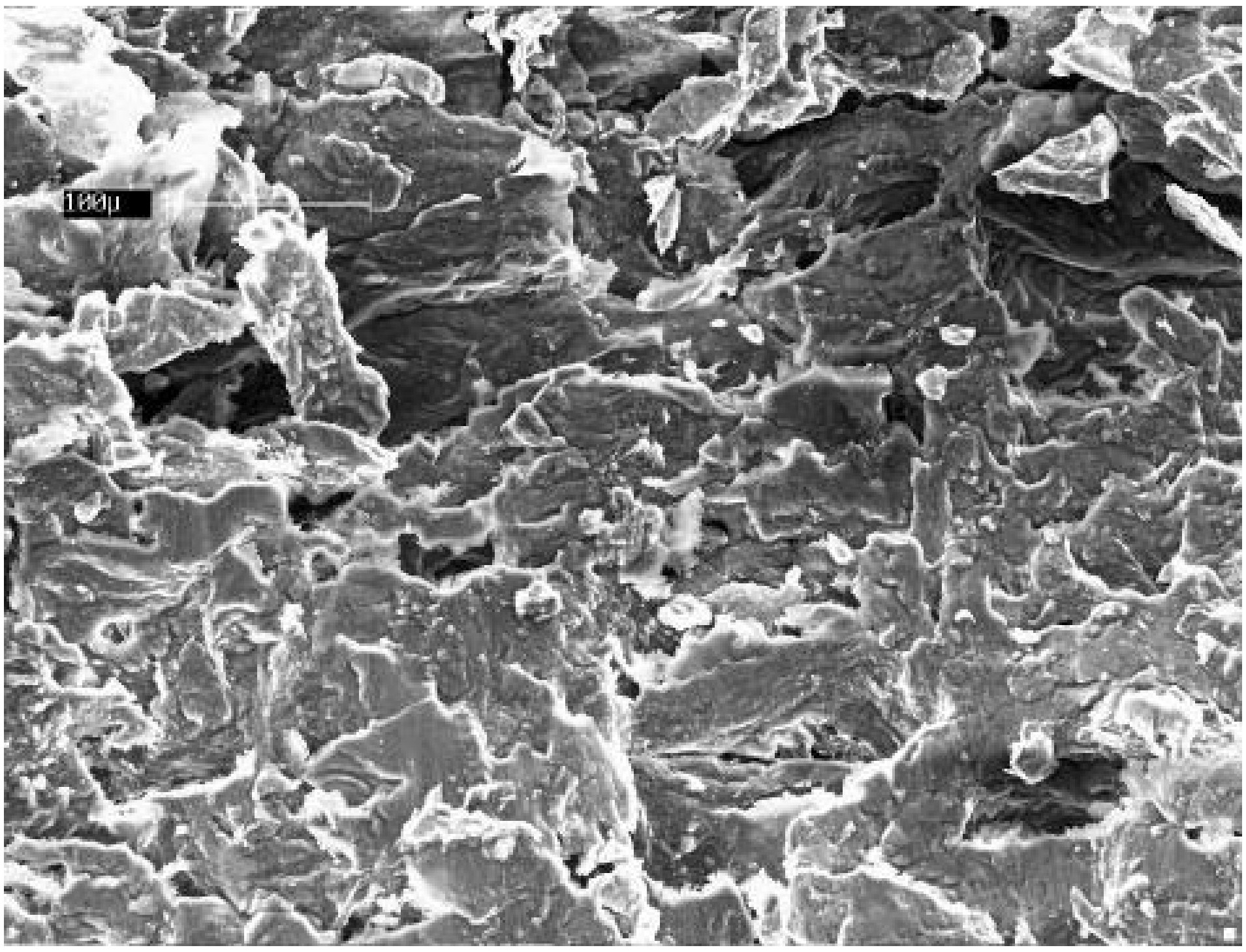

Fig. 2.

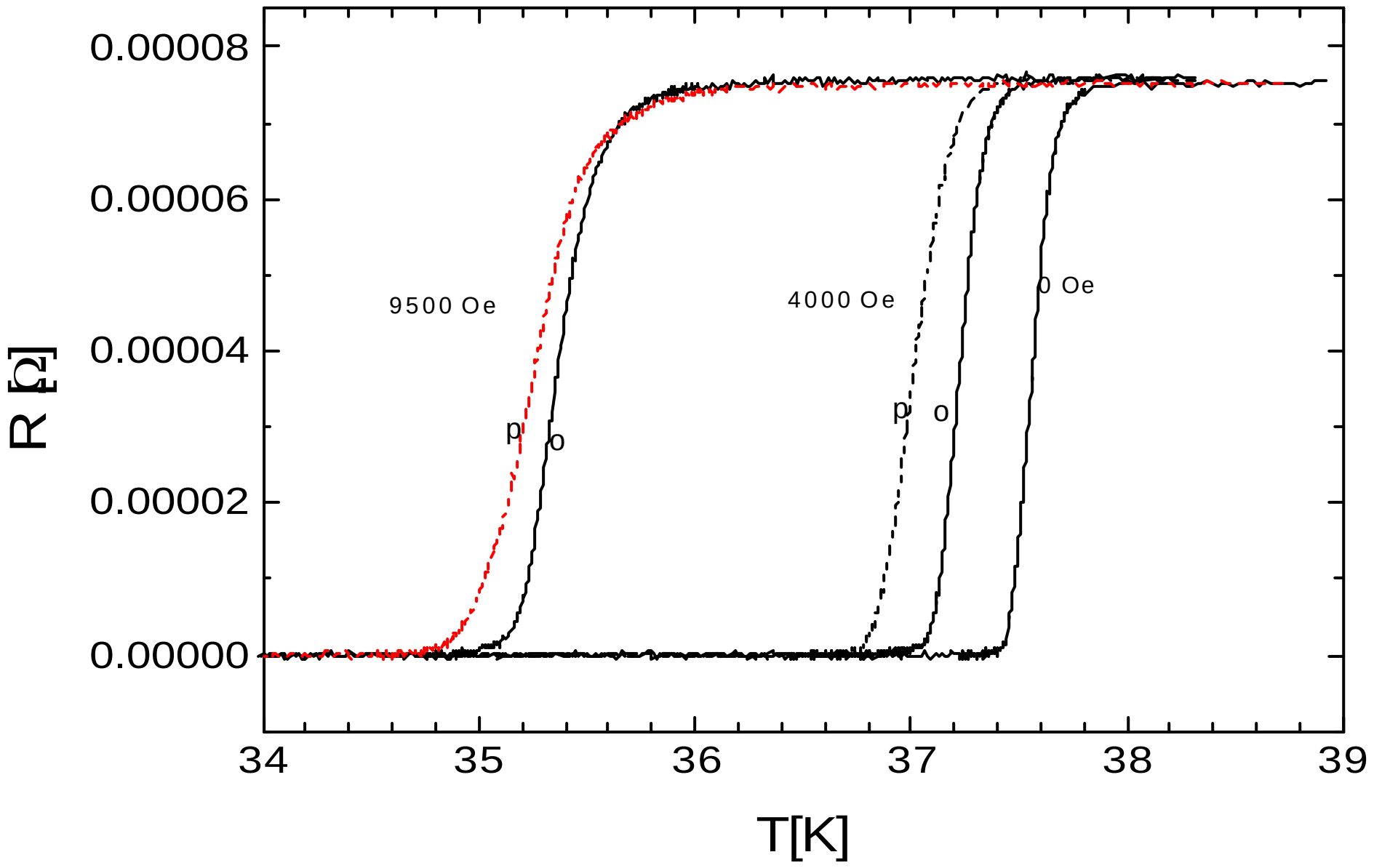

Fig.3.

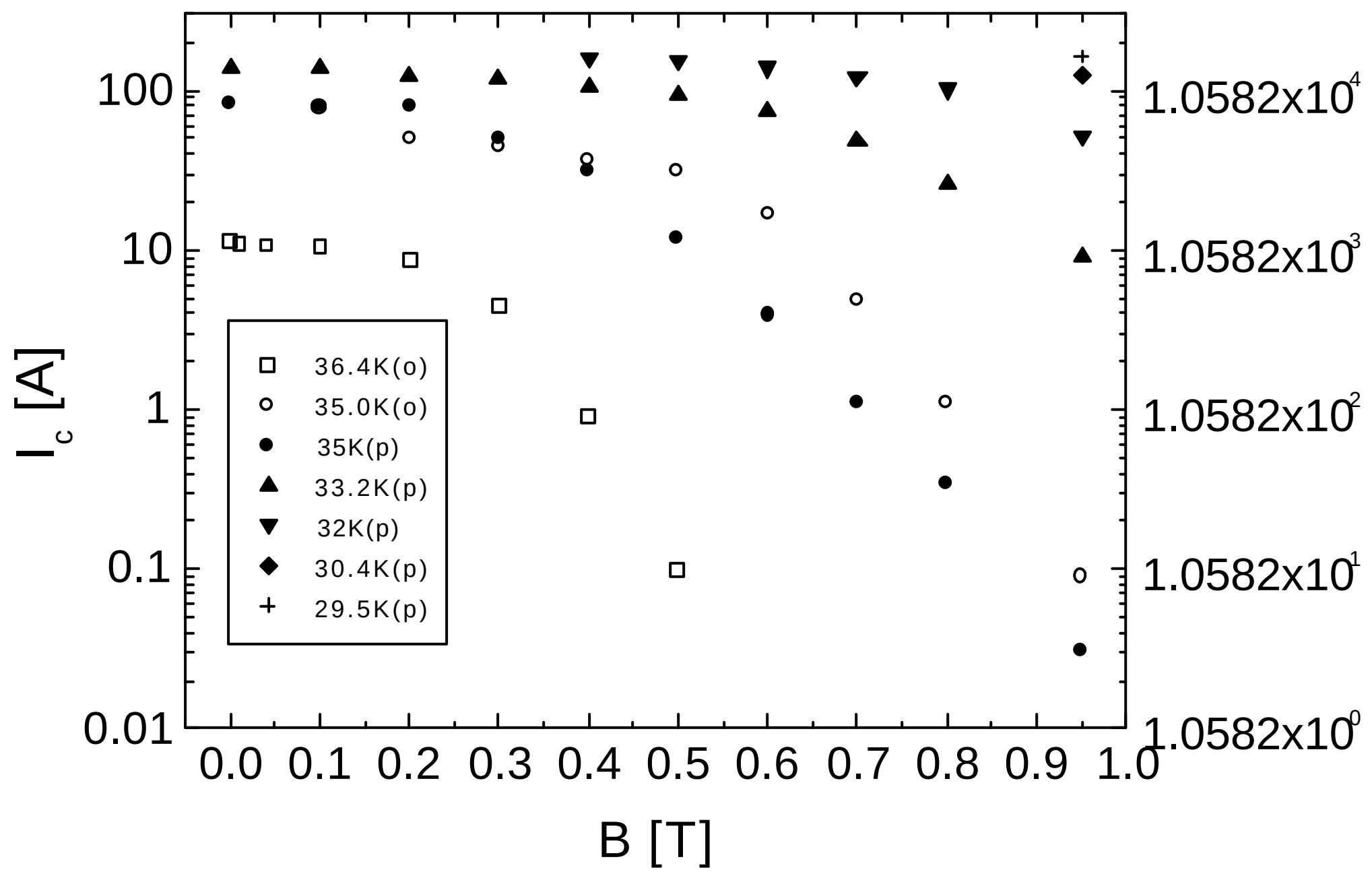
I_c [A]
B [T]
0.01
0.1
1
10
100
0.0 0.1 0.2 0.3 0.4 0.5 0.6 0.7 0.8 0.9 1.0
1.0582x10^0
1.0582x10^1
1.0582x10^2
1.0582x10^3
1.0582x10^4
36.4K(o)
35.0K(o)
35K(p)
33.2K(p)
32K(p)
30.4K(p)
29.5K(p)

Fig. 4.